\documentclass[iop]{emulateapj}

\usepackage{color}
\usepackage{amsmath}
\usepackage{multirow,bigdelim} 

\newcommand       \mum          {\,{\rm \mu m}}

\makeatletter
\def\ExtendSymbol#1#2#3#4#5{\ext@arrow 0099{\arrowfill@#1#2#3}{#4}{#5}}
\def\RightExtendSymbol#1#2#3#4#5{\ext@arrow 0359{\arrowfill@#1#2#3}{#4}{#5}}
\def\LeftExtendSymbol#1#2#3#4#5{\ext@arrow 6095{\arrowfill@#1#2#3}{#4}{#5}}
\makeatother

\begin{document}

\title{Evidence for a Snow Line Beyond the Transitional Radius in the TW Hya Protoplanetary Disk}

\author{K. Zhang\altaffilmark{1}, K.~M. Pontoppidan \altaffilmark{2}, 
C. Salyk\altaffilmark{3}, and  G.~A. Blake\altaffilmark{4}}
\altaffiltext{1}{Division of Physics, Mathematics \& Astronomy, MC 150-21, California Institute of Technology, Pasadena, CA 91125, USA}
\altaffiltext{2}{Space Telescope Science Institute, Baltimore, MD 21218, USA}
\altaffiltext{3}{National Optical Astronomy Observatory, 950 N. Cherry Ave., Tucson, AZ 85719, USA}
\altaffiltext{4}{Division of Geological \& Planetary Sciences, MC 150-21, California Institute of Technology, Pasadena, CA 91125, USA}

\email{kzhang@caltech.edu}

\begin{abstract}

We present an observational reconstruction of the radial 
water vapor content near the surface of the TW Hya transitional protoplanetary disk, and report the first localization of the snow line during this phase of disk evolution. The observations are comprised of \textit{Spitzer}-IRS, \textit{Herschel}-PACS, and \textit{Herschel}-HIFI archival spectra.
The abundance structure is retrieved by fitting a two-dimensional disk model to the available star$+$disk photometry and all observed H$_2$O lines, using a simple step-function parameterization of the water vapor content  near the disk surface. We find that water vapor is abundant ($\sim 10^{-4}$ per H$_2$) in a narrow ring, located at the disk transition radius some 4\,AU from the central star, but drops rapidly  by several orders of magnitude beyond 4.2\,AU over a scale length of no more than 0.5\,AU. The inner disk (0.5-4\,AU)  is also dry, with an upper limit on the vertically averaged water abundance of $10^{-6}$ per H$_2$.  The  water vapor peak occurs at a radius significantly more distant than that expected for a passive continuous disk around a 0.6\,$M_{\odot}$ star, representing a volatile distribution in the TW Hya disk that bears strong similarities to that of the solar system. This is observational evidence for a snow line that moves outward with time in passive disks, with a dry inner disk that results either from gas giant formation or gas dissipation  and a significant ice reservoir at large radii. The amount of water present near the snow line is sufficient to potentially catalyze the (further) formation of planetesimals and planets at distances beyond a few AU.

\end{abstract}

\section{Introduction}

Giant planets and planetesimals form in the chemically and dynamically active environments in the inner zone of gas-rich protoplanetary disks \citep{Pollack96, Armitage11}. Processes taking place during this critical stage in planetary evolution determine many of the parameters of the final planetary system, including the mass distribution of giant planets, as well as the chemical composition of terrestrial planets and moons. 

Indeed, some of the most important features of mature planetary systems may be dictated by the behavior of the water content of protoplanetary disks. Such processes include the formation of a strong radial dependence of the ice abundance in planetesimals -- the \textit{snow line} \citep{Hayashi81}. The deficit of condensible water inside the snow line leads to a need for radial mixing of icy bodies during the later gas-poor stages in the evolution of the planetary system, in particular to explain the existence of water on the Earth \citep{Raymond04}. Ice mantles on dust beyond the snow line allow grains to stick at higher collisional velocities allowing for efficient coagulation well beyond the limits of refractory grain growth alone \citep{Blum08}. The combined mass of the solid component in protoplanetary disks is likely dominated by volatile, but condensible, species, the most abundant of which is water \citep{Lecar06}. 

Thus, core accretion models that rely on the availability of large concentrations of solid material suggest that giant planets will initially be found beyond the snow line, as their formation requires the existence of a massive ice reservoir \citep{Kennedy08, Dodson-Robinson09}. The added solid density offered by condensible water may also offer a way out of the ``meter-barrier'' for planetesimal formation \citep{Weidenschilling97}, in which centimeter to meter size icy bodies migrate inwards and are lost to the central star on time scales much shorter than those of steady state grain aggregation \citep{Birnstiel09}: Beyond a certain mass density of solids, local concentrations of boulder-sized particles may overcome the effects of turbulence acting to disperse them and will gravitationally contract to form Ceres-mass planetesimals \citep{Johansen07, Johansen09}.

Given the central role that water plays in the formation of planetary systems, its distribution and abundance in protoplanetary disks has been the subject of intense theoretical study \citep[e.g.,][]{Ciesla06, Garaud07, Kretke07}. Yet, the only source of observational constraints on the initial  radial distribution of water in planet-forming regions has been the distribution of ice in the current Solar System \citep{Hayashi81}, the remaining bodies of which represent only a small part of the story. Apart from the fact that the solar system water distribution may be unique, theoretical study shows that the initial distribution of ices in protoplanetary disks evolves as does the disk temperature distribution \citep[e.g.][]{Garaud07}, and that in later stages it is confounded by the dynamical interactions between planets and the planetesimal swarm \citep{Gomes05}. The final distribution of water in a planetary system is thus an obscured tracer of initial conditions; and to understand the role of water and other volatiles in the formation of planets, we should measure their distribution before, during and after planet(esimal) formation.  

The most direct way to trace the water abundance distribution in protoplanetary disks is to image line emission from water vapor. However, due to the strong radial gradient of gas temperature in disks, each line of a given excitation energy will only trace a limited areal extent of the disk. For instance, high-J rovibrational transitions in the mid-infrared waveband are sensitive to the warmer ($\gtrsim 300$\,K) and inner parts of the disk surface ($\lesssim 10$\,AU). Far-infrared transitions arise from cooler gas ($\lesssim 300\,$K) in the outer ($\gtrsim 10\,$AU)  and deeperregions of the disk. In order to measure the total water vapor content in the disk surface, it is necessary to observe lines across states of widely varying excitation energies. 

That line emission from water vapor is common in disks is now well established thanks to observations with the \textit{Spitzer} IRS instrument, which has detected a forest of emission lines in the 10--35\,$\mum$ range. \cite{Carr08} and \cite{Salyk08} first reported a large number of rotational lines due to warm water in AA Tau, AS 205N, and DR Tau, while \cite{Pontoppidan10a} report a water emission detection rate of $\sim$50\%~ in their \textit{Spitzer} survey of 46 disks surrounding late-type (M-G) stars. With upper energy levels of $E_{up} \sim $ 1000 $-$ 3000\,K, these water lines arise from the inner few AU region of the disks. Detailed radiative transfer models show that the surface water vapor abundance may need to be truncated beyond $\sim$1\,AU for a typical classical T Tauri star (cTTs) disk in order to match the general line ratios over 10--35$\mum$ \citep{Meijerink09}. Follow-up ground-based spectra of the brightest targets have constrained the gas kinematics, and confirmed that the water vapor  resides inside the snow-line \citep{Pontoppidan10b}. 

Recently, complementary spectral tracers of water vapor in the {\it outer} disk have become accessible via \textit{Herschel} Space Observatory PACS and HIFI observations. \citet{Riviere12} describe the discovery of warm water emission at 63.3$\mum$ in 8 out of  their 68 T Tauri disk sources, and tentative detections of water transitions with similar excitation energies have also been reported for the Herbig Ae/Be star HD 163296 \citep{Meeus12, Fedele12}. The first sensitive search for ground-state emission lines of cold water vapor in DM Tau indicated that the  vertically averaged water vapor abundance in the outer disk of this source is extremely low, $<10^{-10}$ per hydrogen molecule \citep{Bergin10}. Finally, two ground state water emission lines in TW Hya were detected with HIFI \citep{Hogerheijde11}.Again, the vertically averaged water vapor abundance is $_<\atop{^\sim}$$10^{-10}$ per hydrogen, while the estimated peak water vapor abundance is closer to $\sim$10$^{-8} - 10^{-7}$ in the near surface layers of the disk. These are values that can be produced by UV photodesorption from icy dust grains. In this interpretation, the low excitation water lines trace a large, but otherwise unseen, reservoir of ice in the outer disk. 

Here we present an observationally-based method for reconstructing the water vapor  column density  profile in the surface of a protoplanetary disk, based on multi-wavelength, multi-instrument IR/submillimeter spectra, and apply the method to the TW Hya transitional disk. We chose TW Hya because of the availability of deep archival \textit{Spitzer} and \textit{Herschel} water spectroscopy and the existence of detailed structural models. Further, we report on the spectroscopic identification of warm water vapor emission at 20--35\,$\mum$, a detection that permits, in combination with the \textit{Herschel} data, the first localization of the snow line in a transitional disk. 

The outline of the paper is as follows:  In Section \ref{sec:data}, we describe the \textit{Spitzer} and \textit{Herschel} data. In \S \ref{sec:dustmodel}, we describe the structural gas/dust model for the TW Hya disk. Section \ref{sec:radtrans} applies a two-dimensional line radiative transfer model to the model disk structure to retrieve a radial abundance water profile based on the full spectroscopic dataset. The results are discussed in \S \ref{sec:results}.

\section{Data reduction}
\label{sec:data}
The \textit{Spitzer} IRS 10$-$35\,$\mum$ high resolution mode (R$\sim$600) spectra of TW Hya were acquired as part of a survey program of transitional disks (PID 30300), with J. Najita as PI. The observations were done in two epochs using different background observation strategies, for an observation log see \citet{Najita10}. In the first epoch, TW Hya was observed on source (AOR 18017792), followed by north and south off-set observations (AORs 18018048, 18018304).  In the second epoch (AOR 24402944), TW Hya was observed using a fixed cluster-offsets mode, such that the background scans were observed in the same sequence.  All of the datasets were extracted from the \textit{Spitzer} archive and reduced using the Caltech High-resolution IRS Pipeline (CHIP) described in \cite{Pontoppidan10a}, which takes full advantage of the existence of redundant background observations. CHIP implements a data reduction scheme similar to that developed by \citet{Carr08}.

\begin{figure*}[]
\includegraphics[width=16.5cm]{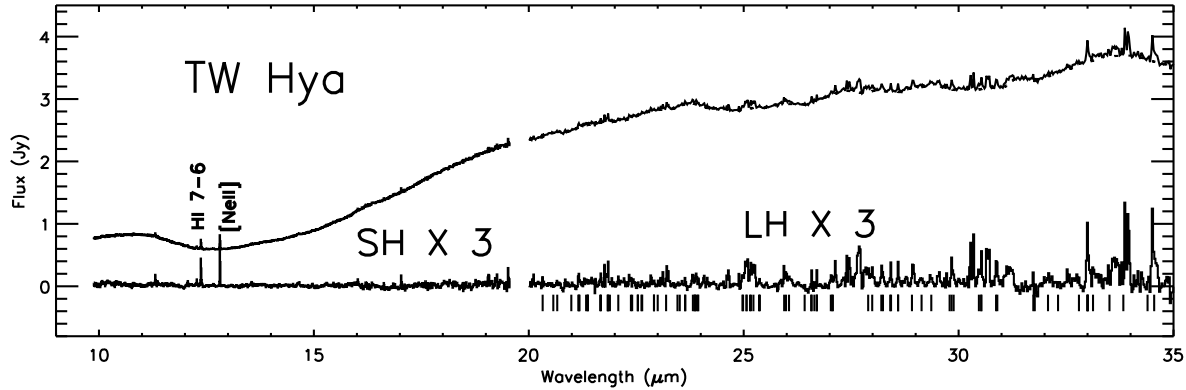}

\caption{ The \textit{Spitzer} IRS high resolution (R=600) spectrum of TW Hya from 10-35\,$\mum$; the trace at bottom presents the data after the subtraction of a spline continuum fit (and scaled by a factor of 3 to show the emission lines more clearly). No water emission is detected in the 10-20\,$\mum$ SH modules \citep{Najita10}, but the 20-35\,$\mum$ LH spectrum shows clear evidence for water vapor. The short vertical ticks indicate water vapor rest wavelengths (HITRAN 2008 database) for lines with intensities $>$5$\times10^{-22}$\,cm$^{-1}$/(molecule cm$^{-2}$). }
\label{fig:spitzer_all}
\end{figure*}

In an extensive analysis of the 10-20\,$\mu$m Short-Hi (SH) spectrum, \citet{Najita10} report the detection of excited OH emission lines and a host of other molecular features from TW Hya, but no water emission above a lower flux limit of $\sim$10\,mJy. As stressed by \citet{Najita10}, the lack of detectable water emission from the highly excited water lines below 20\,$\mu$m does not preclude the presence of cooler water vapor, and in a follow-up paper (in prep), these authors interpret the SH and Long-Hi (LH) data together. Our independent reduction of the deep \textit{Spitzer} IRS observations at longer wavelengths displays numerous water emission lines in the 20$-$35\,$\mum$ LH module. The strongest transition at 33\,$\mum$ can be clearly seen in all LH epochs. The spectrum from the fixed cluster-offsets AOR has the highest signal-to-noise ratio (SNR), and our analysis is therefore based on this spectrum (presented in Figure \ref{fig:spitzer_all}). The SH spectrum is scaled by a factor of 1.25 to match the flux of LH spectrum and IRAS 12\,$\mum$ photometry data.

The \textit{Herschel} HIFI line fluxes for the ground state ortho and para water transitions are taken from \citet{Hogerheijde11}. Somewhat higher excitation lines are probed by the PACS instrument, and narrow-range line spectra of TW Hya from 63$-$180\,$\mum$ were acquired as part of the public \textit{Herschel} Science Demonstration Phase program (observation ID 1342187238). The PACS data were reduced using the \textit{Herschel} interactive processing environment (HIPE  v.6.0), up to level 2. After level 1 processing, spectra for the two nod positions were extracted separately from the central spaxel. The spectra of the two nod positions were uniformly rebinned, using an over-sampling factor of 2 and an up-sampling factor of 2, before co-adding to produce the final result. Linear baselines were fitted to the local continua and integrated line fluxes (see Table~\ref{tab:HIFI}) were obtained using Gaussian fits. While no water emission is formally detected with PACS at 3$\sigma$, there are indications of water lines consistent with our water abundance model predictions; see Section \ref{sec:radtrans} for further discussion.

\section{The TW Hya disk structure} 
\label{sec:dustmodel}
The accuracy of retrieved chemical abundances in protoplanetary disks is coupled to how well the two- or three-dimensional gas kinetic temperature and volume density structures are known. Our model development consists of three stages: First, we construct an axisymmetric model of the dust component of the disk, based on fits to the continuum spectral energy distribution (SED), using the RADMC code\citep{Dullemond04}. We then assume that the gas density distribution follows that of the small dust grain population --- a reasonable assumption since the small grains carry most of the disk opacity and are dynamically well-coupled to the disk gas. However, we do allow thermal decoupling of the gas from the dust near the disk surface, in the form of a constant temperature offset constrained using the CO rovibrational emission spectrum near 4.7\,$\mu$m. Finally, we constrain the water vapor radial column density distribution (and thus abundance) at all disk radii via fits to the multi-wavelength water spectra acquired for TW Hya.  

\subsection{The dust temperature and density structure}

Dust, as the dominant source of disk continuum opacity, determines the disk SED shape. At a distance of only $\sim51\pm 4$\,pc \citep{Mamajek05}, TW Hya is one of the best studied protoplanetary disks, with extensive photometric data available from UV to cm wavelengths. \citet{Calvet02} developed a physically self-consistent dust structure model for TW Hya. They found that the disk is vertically optically thin within 4\,AU, corresponding to a depletion of small dust grains by orders of magnitude at these radii. This inner zone of dust depletion has been confirmed by subsequent observations \citep {Eisner06, Hughes07}. A small amount of dust in  the inner disk is needed to explain the 10--25\,$\mu$m amorphous silicate features \citep{Calvet02}, and accretion onto the star continues.  The existence of hot gas in the innermost disk has been further confirmed by the detection of CO $\Delta v$=1 rovibrational emission near 4.7\,$\mu$m \citep{Rettig04, Salyk07, Salyk09} and via Spectro-Astrometric (SA) observations of these same lines \citep{Pontoppidan08}. It is not yet clear whether the gas content has been depleted to the same degree as the dust in the inner disk, although \citet{Gorti11} estimated enhancements of 5-50 in gas/dust in in the inner disk.

\begin{figure*}
\includegraphics[width=16.5cm]{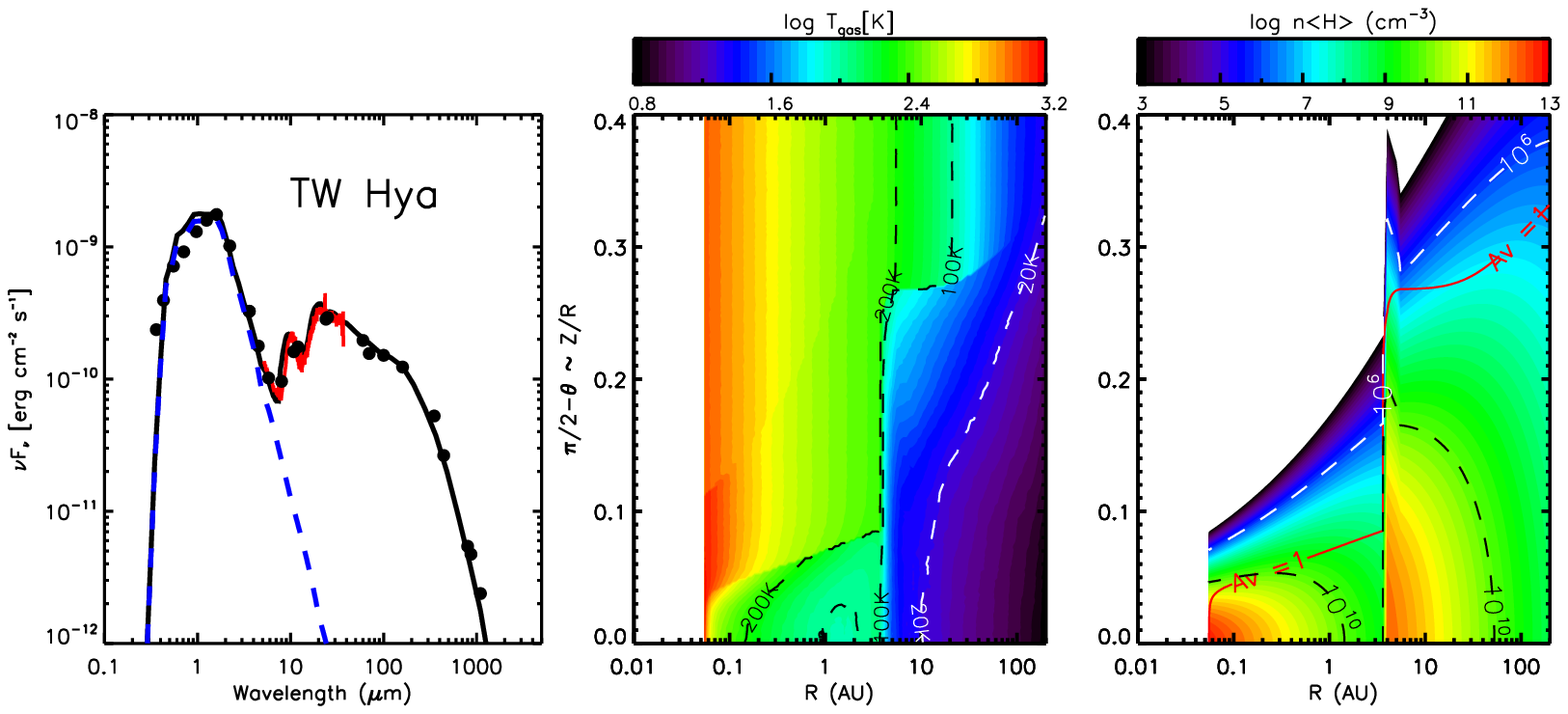}

\caption{Left panel: A 2-D model fit to the TW Hya SED using RADMC and the parameters in Table \ref{tab:para}. The stellar spectrum (peaking near 1 $\mu$m) is generated from a Kurucz model with T$_{\rm eff}$ = 4000\,K. The photometric points are taken from \citet{Rucinski83}, 2MASS, \citet{Low05}, and \cite{Weintraub89}, while the 5--35\,$\mum$ $Spitzer$ IRS spectrum is that displayed in Figure \ref{fig:spitzer_all}. Middle panel: Fiducial gas temperature profile. Right panel:  Fiducial gas density profile, with the $A_v$=1 height depicted by the solid line. }
\label{fig:SED}
\end{figure*}

We adopt M$_{\star}$ = 0.6\,M$_\odot$, R = 1\,M$_\odot$ and T$_{\rm eff}$ = 4000\,K for the stellar mass, stellar radius and effective temperature \citep{Webb99}. The inclination angle of the disk axis to the line-of-sight is fixed to $i$ = 7$^{\circ}$ \citep{Qi04}.  The standard gas-to-dust ratio of 100 is used to estimate the total disk mass. 

The dust structure model adopted is similar to  \citet{Andrews12}: an optically thin inner disk between the sublimation radius ($r_{\rm sub}$) and the cavity edge ($r_{\rm cav}$), a cavity ``wall" and an optically thick disk.  The cavity ``wall" component is needed to reproduce the spectral shape of TW Hya from 10$-$30$\mum$ \citep{Uchida04}. The sublimation radius was set to $r_{\rm sub}$ = 0.055\,AU, the location where dust temperature reaches 1400\,K. The cavity radius is fixed as $r_{\rm cav}$ = 4\,AU, in accord with previous near-IR interferometry \citep {Eisner06, Hughes07, Akeson11}.  We adopt the power-law density profile of \citet{Andrews12} for the outer disk, which is based on 870 $\mum$ continuum interferometric imaging.  A summary of the dust structure model parameters can be seen in Table 1. 

Following \citet{Pollack94} and \citet{DAlessio01}, we use a dust mixture of astronomical silicates, organics and water ice.  For the optically thin inner disk, pure glassy silicate is used to match the strong silicate bands at 10 and 18\,$\mum$. The dust size distribution is taken to be $n(a) \propto a^{-s}$ with $s$ = 3.5 between  $a_{\rm min}=0.9$ and  $a_{\rm max}=2.0$ $\mu$m \citep{Calvet02}.  In the cavity wall, we use a dust size range between 0.005 and 1\,$\mum$; outside the cavity wall, 95\% (by mass) of the dust has a size distribution extending from 0.005\,$\mum$ to 1\,mm, with the remaining 5\% contained in grains from 0.005 -- 1\,$\mum$. Total dust masses in the optically thin and thick parts of the disk are $2.77\times 10^{-9} $ and  $4.4\times 10^{-4} $ M$_\odot$.  The SED fit and our fiducial gas temperature and density profiles, discussed further below, are presented in Figure \ref{fig:SED}.

\begin{deluxetable}{llll}
\tablecolumns{4}
\tablewidth{0pt} 
\tablecaption{Disk parameters for the dust model}
\tablehead{\colhead{Parameter}  & \colhead{Inner disk}&  \colhead{Cavity wall}& \colhead{Outer disk} 
}
\startdata
$r$ (AU)  & 0.055 $-$ 4  &4 $-$5 & 5$-$196\\
$p$ & 0.8 & 0.0 & 1.0\\
$\gamma$  & 0.3 & -0.25 &0.2
\enddata
\tablecomments{Here $\gamma$ is the index that describes the vertical 
scale height $z_d$: $z_d = z_0(r/r_0)^{1+\gamma}$ where $z_0$ = 0.7 at $r_0 = 10$\,AU; $p$ is the index of the surface density 
distribution: $\Sigma_d = \Sigma_0(r/r_0)^{-p}$. }

\label{tab:para}
\end{deluxetable}

\subsection{The gas density and temperature}

The gas/dust ratio in disks evolves due to grain growth and transport processes \citep{Birnstiel09}, and likely has significant radial and vertical structure. For simplicity, the gas density is here assumed to follow that of the dust, with a constant gas/dust mass ratio throughout the disk. For TW Hya, the estimated global gas/dust ratio varies from 2.6 to 100 \citep{Thi10, Gorti11}. The recent detection of the HD J = 1-0 line from TW Hya has offered an independent and robust estimation of the gas mass \citep{Bergin12}, which indicates a globally averaged gas/dust ratio close to 100, a value we adopt here. Significant dust vertical settling will produce an enhanced gas-to-dust ratio in the upper layers of the disk. Such settling should have little impact on the longest wavelength water transitions studied here, or on the strongest, optically thick lines traced by the Spitzer IRS. By creating a larger column of gas above the $\tau_{dust}=1$ surface, the absolute fractional abundance of water derived by a well mixed LTE model likely provides an upper bound, but the inferred radial \textit{structure} in the water vapor column density should be fairly robust against changes to the dust distribution.

Above a certain altitude in the disk, where the environment becomes exposed to the ambient radiation field, the gas can also be thermally decoupled from the dust. The gas will adopt a temperature profile that is a balance among heating processes, such as mechanical heating via accretion or that driven by photodissociation or the photoelectric electric effect, and cooling rates driven by atomic, molecular, and dust grain emission \citep{Glassgold04, Kamp04}. General gas temperature profiles can be estimated using detailed thermo-chemical models (e.g. \citealp{Woitke09, Najita11}). However, such profiles can be highly dependent on input assumptions, and interdependencies between model parameters and observables can be obscure. Here, we simplify the process by deriving the vertical gas temperature structure in the inner disks using the rotational ladder from M-band CO P-branch $v=1-0$ emission lines. Due to its (photo)chemical robustness, the abundance of CO is predicted to be fairly constant at $n_{CO}/n_{H_2} \approx 1.2\times10^{-4}$ in regions warmer than 20 K \citep{Aikawa96}. A high CO abundance is expected to persist even for a depleted inner disk, such as that of TW Hya. Indeed, in the chemical model of \citet{Najita11}, the abundance of CO  rises to $\sim 10^{-4}$ once the  vertical column density of H$_2$ reaches 10$^{21}$ cm$^{-2}$ for radii beyond 0.25\,AU (physical densities are $>$10$^9$ cm$^{-3}$).  

Since the heating of the inner disk gas is driven by X-ray and FUV photons from the central star and stellar accretion flow \citep{Kamp04, Gorti11}, we assume that gas and dust temperature become decoupled in regions where the radial optical depth for visible photons is less than unity. That is:
 
\begin{equation}
\label{eq:tgas}
T_{\rm gas}=\left\{\begin{array}{ll}
T_d, & A_v >1 \\
T_d + \delta T , & A_v \leqslant 1
\end{array}\right.,
\end{equation}

\noindent where T$_d$ is dust temperature in the disk at (r, $\theta$) in spherical coordinates, $A_v$ is the visual extinction along the radial direction $\theta$, and $\delta$T is a free parameter that describes the gas-dust temperature difference, to be determined through fits to observational data. 

\begin{figure}
\includegraphics[width=6.5cm, angle=90]{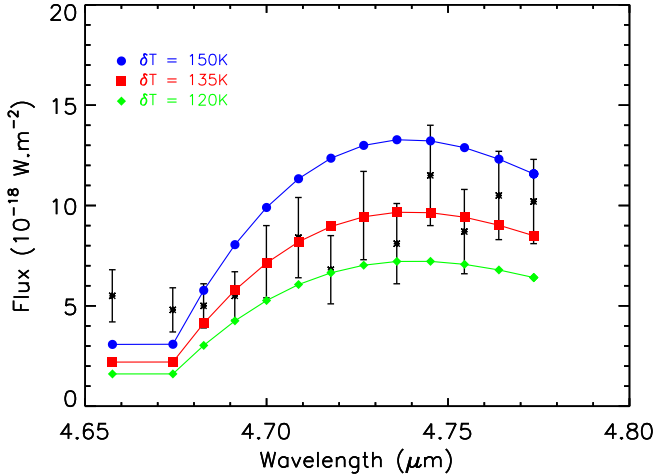}
\vspace{0.0001in}
\caption{Model CO rovibrational fluxes from decoupled gas/dust temperature models compared with the observational data (depicted as crosses with error bars, from \citealt{Salyk07}). The models show three different values of $\delta$T, the temperature difference between gas and dust, with values of 120 (diamonds), 135 (squares) and 150\,K (circles).  }
\label{fig:CO}
\end{figure}

The $v$=1-0 P(1-12) CO line fluxes at 4.7\,$\mum$ are taken from \citet{Salyk07}, while the model spectra are generated by the raytracing code RADLite \citep{Pontoppidan09}. The lines are $\sim$7.5\,km/s wide (FWHM) \citep{Pontoppidan08}, consistent with the nearly face-on orientation of TW Hya. The inner edge of the CO emission in the inner disk is set at r$_{in}$ = 0.11\,AU, based on SA imaging of the 4.7\,$\mum$ line emission \citep{Pontoppidan08}. The outer edge of the CO-emitting zone is not well constrained, and is set to 4\,AU, the dust transition radius.  The model is relatively insensitive to the size of the outer edge, since the majority of the emission is produced at radii $\ll4$\,AU.  
 
The best fit has $\delta$T=135$\pm$15\,K (Figure \ref{fig:CO}). From M-band rotation diagram fits, \citet{Salyk07} derive N$_{\rm CO}$ = 2.2 $\times$ 10$^{18}$ cm$^{-2}$, corresponding to a vertical total gas column density of N$_{\rm H}$$\sim$10$^{22}$\,cm$^{-2}$. The 135\,K gas-to-dust temperature difference is consistent with the detailed gas temperature balance model of \citet{Najita11}, whose treatment yields $\delta$T$\sim$100 K at r=0.25\,AU for an integrated column density of N$_{\rm H}$ = 10$^{22}$\,cm$^{-2}$. RADLite simulations predict a line width (prior to instrument profile convolution) of FWHM$\sim$6.8-9.6\,km/s for $v_{\rm turb}\sim 0.05\,v_{\rm Kepler}$.  This is somewhat larger than observed 7.5\,km/s linewidths, but this difference may be accounted for by the difference in disk inclination relative to that ($i=4^\circ$) derived by \citet{Pontoppidan08}.

The best fit $\delta$T of 135\,K applies only to the inner disk radii probed by CO, but similar physics will decouple the gas and dust temperatures near the disk surface at larger radial distances. To model this decoupling, we adopt $T_g = T_d + \delta T\times e^{-r/50\, \mathrm{AU}} $ for the gas temperature at large scale heights, a parameterization that matches broadly the models of \citet{Thi10} for TW Hya. Again, the gas density structure follows that of the dust.

We stress that such gas/dust thermal decoupling should have only a modest impact on a principle molecular mapping result presented here, namely the significant drop in the water vapor column density beyond the snow line. For the highest excitation lines measured by \textit{Herschel} (and \textit{Spitzer}) that trace the photon-dominated, and thus heated, layers in the inner disk and cavity wall, gas densities are at least $\sim$10$^9$ H$_2$/cm$^3$. Under such conditions the simulations of \citet{Meijerink09} show that the emergent fluxes of the water lines detected here are within factors of two-three of their LTE values. Thus, the water vapor column densities should be reasonably well determined by LTE calculations whose temperature distributions are constrained by the CO M-band observations.

The outer disk is too cold to emit in such high excitation water lines, and as discussed further in Section \ref{sec:results}, the physical density in the $A_v$$\sim$1 layer at radii beyond 5-10 AU is only $\sim$10$^7$ H$_2$/cm$^3$, some 100-1000$\times$ less than the critical densities of the mid- to far-infrared water vapor lines examined here. Only the lowest-excitation water transitions measured by \textit{Herschel} principally trace the outer, deeper disk layers. Here the M-band CO observations are of little use, but the lowest-J pure rotational lines do probe the relevant disk gas. In RADLite simulations of lines up to J = 6-5 we find, in agreement with \citet{Qi04}, that significant CO depletion in the cold disk midplane is needed to match the observed line fluxes. Water can be expected to deplete in a similar manner, with the bulk of the vapor emission arising from an intermediate layer at densities where LTE calculations offer column density estimates good to within an order-of-magnitude \citep{Hogerheijde11}.

\section{Retrieving the radial water vapor profile}
\label{sec:radtrans}
\subsection{The water line model}
Given the gas temperature and density structure for TW Hya, we can now  constrain the water vapor content of the TW Hya disk surface as a function of radius. Our goal is not to provide an exacting description of the volatile abundances versus radius and height, but to determine what radial distribution of water vapor is most consistent with the available data.To this end, we construct a step function in water vapor abundance, with one value, $X_{\rm inner}$, in the inner gas-depleted disk within 4\,AU, another, $X_{\rm ring}$, in a ring starting at 4\,AU and extending to the snow line at $R_{\rm snow}$, and an outer disk abundance $X_{\rm outer}$. That is,
\begin{equation}
\label{eq:stepf}
X_{\rm H_2O}=\left\{\begin{array}{lll}
X_{\rm inner}, & R< 4\, {\rm AU} \\
X_{\rm ring}, & 4\, {\rm AU}\leqslant R\leqslant R_{\rm snow}\\
X_{\rm outer}, &R> R_{\rm snow}
\end{array}\right.
\end{equation}

\noindent  As described below, our calculations assume LTE but do account for line and continuum opacity. The $X_{\rm H_2O}$ values derived thus reflect the water vapor column densities to different depths into the disk. Inside of 4 AU and for the longest wavelength {\it Herschel} lines sensitive to the outermost radii the dust opacity permits the full vertical extent of the disk to be sampled. For the IRS features and the shorter wavelength PACS lines that sample gas near the transition radius, significant dust opacity limits the fitted column densities, and hence the $X_{\rm H_2O}$ values, to the upper layers of the disk. Freeze-out and dust/line opacity greatly limit access to the midplane beyond 4 AU.

Given the wide span in upper level energies of the lines considered, it is generally possible to identify subset of lines that co-vary. Consequently, we can explore one model parameter at a time to derive a best-fit model, illustrated in Figure \ref{fig:column}. As we shall see, the simplest LTE model, namely a constant water abundance throughout the disk, is inconsistent with the data in hand.In Section 4.2.2, we further discuss whether the data support a modification to this basic structure in which the drop in water abundance beyond the snow line instead occurs over some finite, measurable distance.

To render model spectra, we set the level populations to LTE. While the critical densities ($n_{\rm crit} = 10^{10} - 10^{12}$ cm$^{-3}$) of the high energy mid-IR water transitions suggest that a non-LTE treatment is needed, there is currently no robust and fully tested non-LTE framework for the modeling of infrared water lines. That is, a non-LTE calculation is also likely to be inaccurate, given the uncertainty of collisional rates and the complexity of the transition network. Further, previous work suggests that non-LTE effects may alter line fluxes by  factors of only a few for the moderate excitation lines detected here \citep{Meijerink09, Banzatti12}, which will preserve the qualitative aspects of our treatment. In setting level populations to LTE, we make it easier to reproduce our results and to evaluate the validity of our retrievals using more detailed models in the future. 

We assume a subsonic turbulent velocity broadening of $v_{\rm turb}\sim 0.05\, v_{\rm Kepler}$, and a constant ortho-to-para (O/P) water ratio of 3:1, that consistent with the mid- and far-infrared lines \citep{Pontoppidan10a}. The cold gas seen by HIFI has a lower O/P ratio of 0.7 \citep{Hogerheijde11}, but the difference does not substantially affect the derived abundance structure. Model spectra are typically rendered with a grid of 0.3\,km/s, and then convolved with a Gaussian instrument response function to match the observed spectra. The \textit{Herschel} HIFI spectra are rendered with a finer velocity grid of 0.05\,km/s.

\begin{deluxetable*}{lllllll}
\tablecolumns{4}
\tablewidth{0pt} 

\tablecaption{Observed and calculated water line fluxes}
\tablehead{
\colhead{Type}     & \colhead{Transition}  & \colhead{$\lambda(\mum)$}& \colhead{ E$_{up}$ (K)} & A coefficient(s$^{-1}$)&\colhead{ Flux($10^{-19} W/m^2$)} & \colhead{Model($10^{-19} W/m^2)$} 
}
\startdata
  o-H$_2$O   &     11$_{39}$-10$_{010}$   &  17.23   &   2438.8603   &   9.619E-1    & $<$ 163.7&  37.6\\
  p-H$_2$O   &     11$_{29}$-10$_{110}$   &  17.36   &   2432.5234   &   9.493E-1    &  $<$ 617.3& 110.4\\
  o-H$_2$O   &        8$_{36}$-7$_{07}$   &  23.82   &   1447.5970   &   6.065E-1    &  \rdelim\}{6}{1mm}[4669$\pm$364$^a$] &  \multirow{6}{*}{3051}\\
  p-H$_2$O   &        9$_{81}$-8$_{72}$   &  23.86   &   2891.7024   &   3.128E+1    &   \\
  o-H$_2$O   &        9$_{82}$-8$_{71}$   &  23.86   &   2891.7022   &   3.128E+1    &   \\
  o-H$_2$O   &        8$_{45}$-7$_{16}$   &  23.90   &   1615.3501   &   1.034    &   \\
  o-H$_2$O   &      11$_{66}$-10$_{55}$   &  23.93   &   3082.7639   &   1.612E+1    &   \\
  p-H$_2$O   &      11$_{56}$-10$_{47}$   &  23.94   &   2876.1493   &   9.754    &   \\
  p-H$_2$O   &        8$_{53}$-7$_{44}$   &  30.47   &   1807.0023   &   8.923    &    \rdelim\}{3}{1mm}[3750$\pm$317$^a$] &  \multirow{3}{*}{4940}\\
  p-H$_2$O   &        7$_{61}$-6$_{52}$   &  30.53   &   1749.8575   &   1.355E+1    &   \\
  o-H$_2$O   &        7$_{62}$-6$_{51}$   &  30.53   &   1749.8506   &   1.355E+1    &   \\
  o-H$_2$O   &        8$_{54}$-7$_{43}$   &  30.87   &   1805.9308   &   8.687    &    \rdelim\}{2}{1mm}[1883$\pm$287$^a$] &  \multirow{2}{*}{3393}\\
  o-H$_2$O   &        6$_{34}$-5$_{05}$   &  30.90   &    933.7488   &   3.504E-1    &   \\
  o-H$_2$O	 &	  8$_{18}$-7$_{07}$	  &	  63.32	 &	1070.7757	 & 1.730&	  $<$	123.2   &   84.8\\
o-H$_2$O	 &	  7$_{07}$-6$_{16}$	  &	  71.95	  &	843.5459& 1.147	&	  $<$	127.2   &   77.0\\    
p-H$_2$O	 &	  8$_{17}$-8$_{08}$	  &	  72.03	  &	1270.3910	& 3.099E-1&	  $<$	35.4        &   11.6\\
o-H$_2$O	 &	  4$_{23}$-3$_{12}$	  &	  78.74	  	&432.1913	&4.852E-1&	  $<$	39.1        &   76.3\\
p-H$_2$O	 &	  6$_{15}$-5$_{24}$	  &	  78.93	  	&781.1873	&4.501E-1&	  $<$	54.3        &   33.7\\
p-H$_2$O	 &	  3$_{22}$-2$_{11}$	  &	  89.99	  	&296.8471	&3.541E-1&	  $<$	60.9        &   45.8\\
p-H$_2$O	 &	  4$_{13}$-3$_{22}$	  &	  144.52	 &	396.4126&	3.345E-2&    $<$	5.9	      & 8.5 \\
o-H$_2$O	 &	  2$_{12}$-1$_{01}$	  &	  179.53	 &	114.3874&5.617E-2	&    $<$	8.6	      & 10.7\\
p-H$_2$O 	 &		1$_{11}$-0$_{00}$	  &	  269.27	  &	53.4366&1.852E-2	&   3.07$\pm$ 0.19  &	 3.24 \\
o-H$_2$O 	 &		1$_{10}$-1$_{01}$	  &	  539.29	  &	60.9686&	3.477E-3&   3.46$\pm$ 0.15  &	1.89
\enddata
\tablecomments{ a. Water lines are often blended in the $Spitzer$ IRS LH spectrum due to the limited resolution of R$\sim$600.  The fluxes are measured by integrating the spectrum over each distinctive feature/region after a linear continuum subtraction. }
\label{tab:HIFI}

\end{deluxetable*}

\subsection{Best-fitting model}

\begin{figure}
\includegraphics[width=8.5cm]{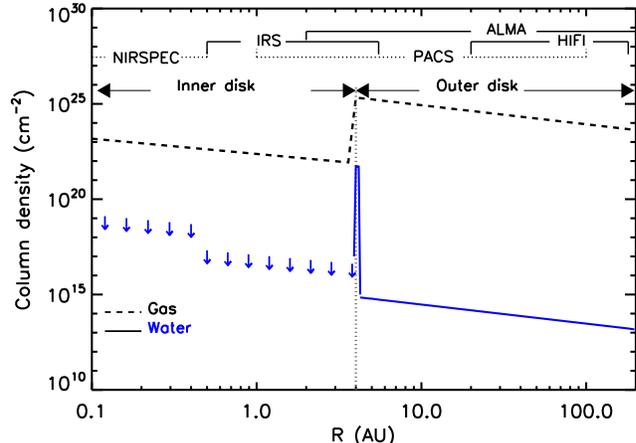}
\caption{The best-fit radial abundance of water vapor in TW Hya, illustrated by the vertically integrated column density profile (solid curve). Within 4 AU, the maximum water abundance is depicted by arrows. The dashed curve is the total gas column density. At the top of the figure, the role of multi-wavelength observations in tracing different radii of the disk are highlighted, along with representative facilities of each wavelength window.  }
\label{fig:column}
\end{figure}

\begin{figure}
\includegraphics[width=6.6cm,angle=90]{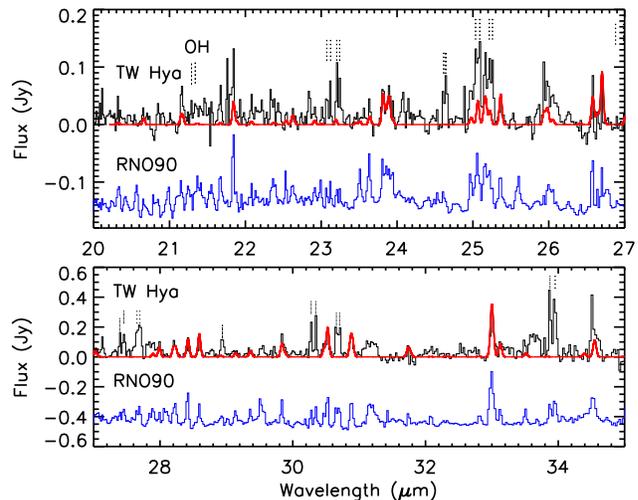}
\caption{Detailed comparisons of the \textit{Spitzer} IRS spectrum (black) with RADLite LTE water models (red). The spectrum of RNO90, one of the strongest water emitters in \cite{Pontoppidan10a}, is plotted at bottom, to guide the eye.  The vertical dashed markers show the location of OH lines not discussed in this paper. }
\label{fig:spitzer_LH}
\end{figure}

The best-fitting radial water column density profile in TW Hya is shown in Figure \ref{fig:column}, whose spectral predictions are compared to the \textit{Spitzer} LH data in Figure \ref{fig:spitzer_LH}. We find that the inner optically thin region of TW Hya is {\it dry}, with upper limits on the vertically integrated fractional water abundance of $X_{\rm inner} < 10^{-6}$. The water detected by \textit{Spitzer} originates in a thin ring near the disk surface, starting at 4\,AU and ending at a snow line just beyond this at $R_{\rm snow}\sim 4.2\,$AU. At larger radii, the fractional water abundance drops abruptly by {\it several orders of magnitude}, to $X_{\rm outer}$ values near 3.5$\times10^{-11}$. 

\begin{center}
\begin{figure*}[]
\includegraphics[width=14cm,angle=90]{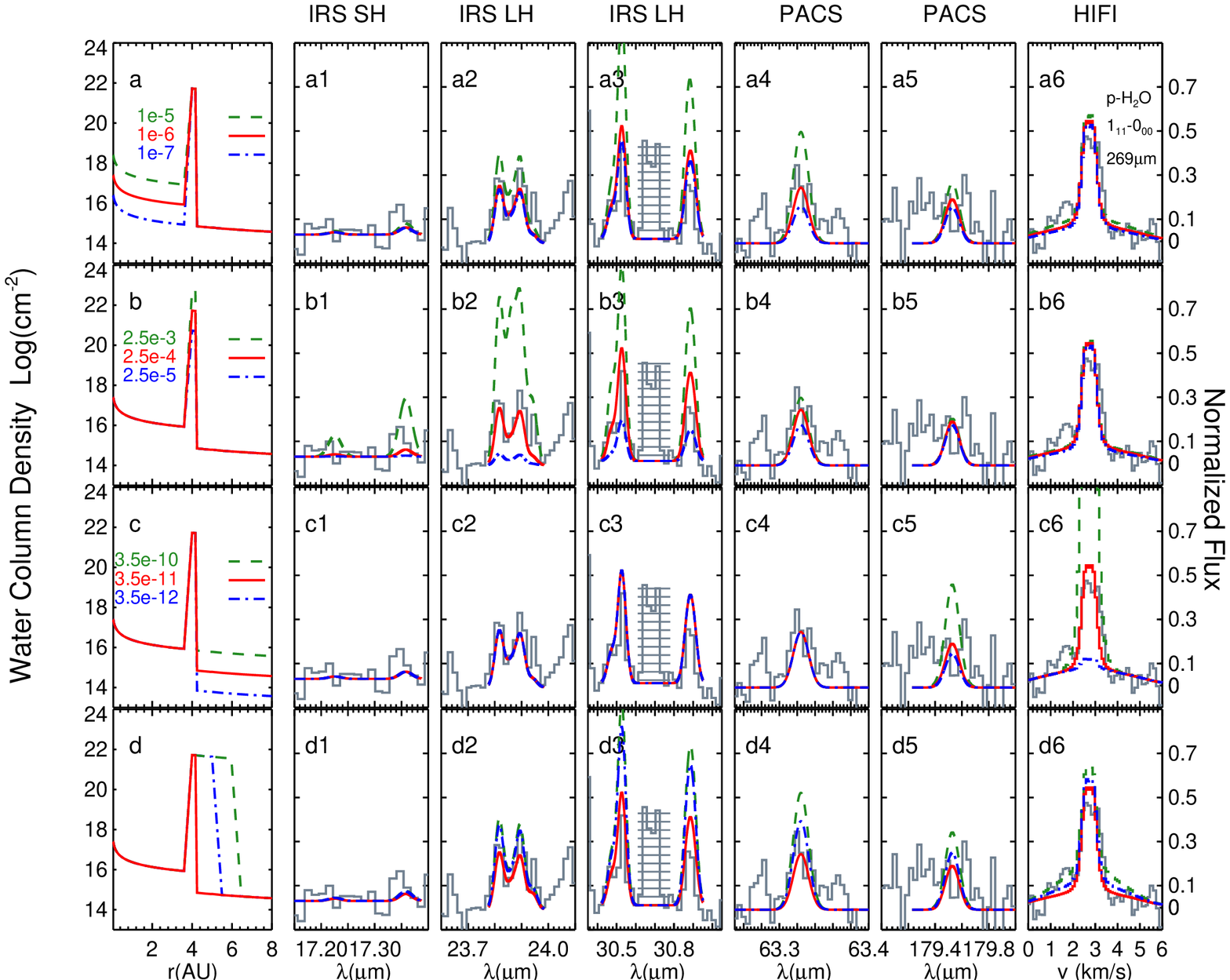}
\caption{
The sensitivity of selected water line fluxes to variations in the water vapor column density distribution. The first column shows the model water vapor distributions. The best-fit model is the solid red curve, while the green and blue curves are models calculated to illustrate the sensitivity of the various lines to changes in the water abundance at different disk radii. The numbers in the first column indicate the fractional (per hydrogen molecule) water abundance in (a) the optically thin inner disk, (b) the ring at the transition radius and (c) the outer disk. The final (d) panel shows the sensitivity of the model to the radius of the snow line.  The following six columns show model line spectra compared to the observed spectra for TW Hya.  The ladder crossed line in the 30$\mum$ column is OH, which is not discussed in this paper. Please see Fig. \ref{fig:spitzer_LH} for additional OH line identifications. }
\label{fig:multi}
\end{figure*}
\end{center}

The uniqueness of this solution is illustrated in Figure \ref{fig:multi}, which shows how variations in the water abundance in different regions of the disk affect different lines. This differential sensitivity is the foundation of our method for retrieving the radial column density, and thus abundance, profile, and is the subject of the following discussion.

\subsubsection{The inner disk}
The inner disk of TW Hya -- radii within 4\,AU -- is partially cleared out, leading to integrated vertical column densities that are more than two orders of magnitudes lower than those typical for cTTs. The non-detection of  water lines between 10 and 20\,$\mu$m, in the \textit{Spitzer} SH range, may be a reflection of this, but even though the column density in the inner disk is dramatically reduced from the extrapolation of the outer disk to smaller radii, the scale height is sufficiently small that the physical densities at the $A_v$$\sim$1 surface exceed 10$^9$ H$_2$/cm$^3$. Thus, non-LTE effects are likely to be fairly unimportant, and we are able to produce meaningful upper limits on the fractional water abundance at $r <$4 AU.

We investigated the effects on the \textit{Spitzer} and \textit{Herschel} lines when varying the inner disk abundance ratios between  $X_{\rm inner} = 10^{-7}$ and $10^{-4}$. Interestingly, we find that inner disk fractional abundances higher than $10^{-6}$ significantly overpredict several water lines in the PACS range, especially the high excitation 63.3$\mum$ lines, while the \textit{Spitzer} SH lines allow somewhat larger abundances. That is, the \textit{Spitzer} SH spectrum tends to be consistent with a model in which the lack of lines below 20\,$\mu$m is explained by the overall low gas mass in the inner disk, and not necessarily a drier disk. It is the addition of the non-detection of PACS lines that require the fractional water abundance to be suppressed in the inner disk. 

\subsubsection{The transition region and the snow line}
Next, we consider the origin of the water lines seen at 20-35\,$\mu$m. Figure \ref{fig:multi} shows that the \textit{Spitzer} LH lines detected are uniquely sensitive to variations in the water vapor content around the transition region. It is clear that a high water vapor column density is required to fit the \textit{Spitzer} spectra, with an emitting area close to $\pi \times$1 AU$^2$. Our fiducial dust/gas model yields $X_{\rm ring}\sim 2.5\times 10^{-4}$ above the $\tau_{\rm dust}$=1 surface (line opacity is also significant). As discussed in \S 5.2, the required column density/abundance can be traded off against excitation temperature to some extent, with higher excitation temperatures corresponding to slightly smaller radii. 

In the context of our transitional disk model, the most likely location for warm water vapor lies at the transition radius, or 4 AU, a prediction that can be tested by measurements of the water (or perhaps OH) emission line profiles using high resolution thermal-infrared spectroscopy. Interestingly, the cavity ``wall'' structure needed to reproduce the 10-30 $\mu$m SED (with a $z/r$ value of 0.3) has a surface area close to that derived from the water emission for an inclination of seven degrees, further reinforcing the likely radial location of the high water vapor abundance.

Further, the PACS and HIFI lines are sensitive to the location of the snow line (the outer edge of the water-rich ring). Consequently, we varied the location of the snow lines between 4.2-6\,AU. In Figure \ref{fig:multi}, it is seen that placing the snow line farther out in the disk over-predicts primarily the PACS lines. Hence, the detection of water lines by \textit{Spitzer} in combination with the non-detection of water lines in PACS waveband provides strong constraints on the location of the surface snow line. 

We explored one more aspect of the water abundance profile around the snow line. Because the water condensation temperature is a function of pressure and because of the strong vertical gradients in the disk, the local water vapor abundance need not precisely be a step function in radius. In particular, can the water lines between 40-150\,$\mu$m be used to constrain how rapidly the water abundance, as inferred from the column density structure, drops beyond the snow line with radius?

\begin{figure*}[!htbp]
\includegraphics[width=15.5cm]{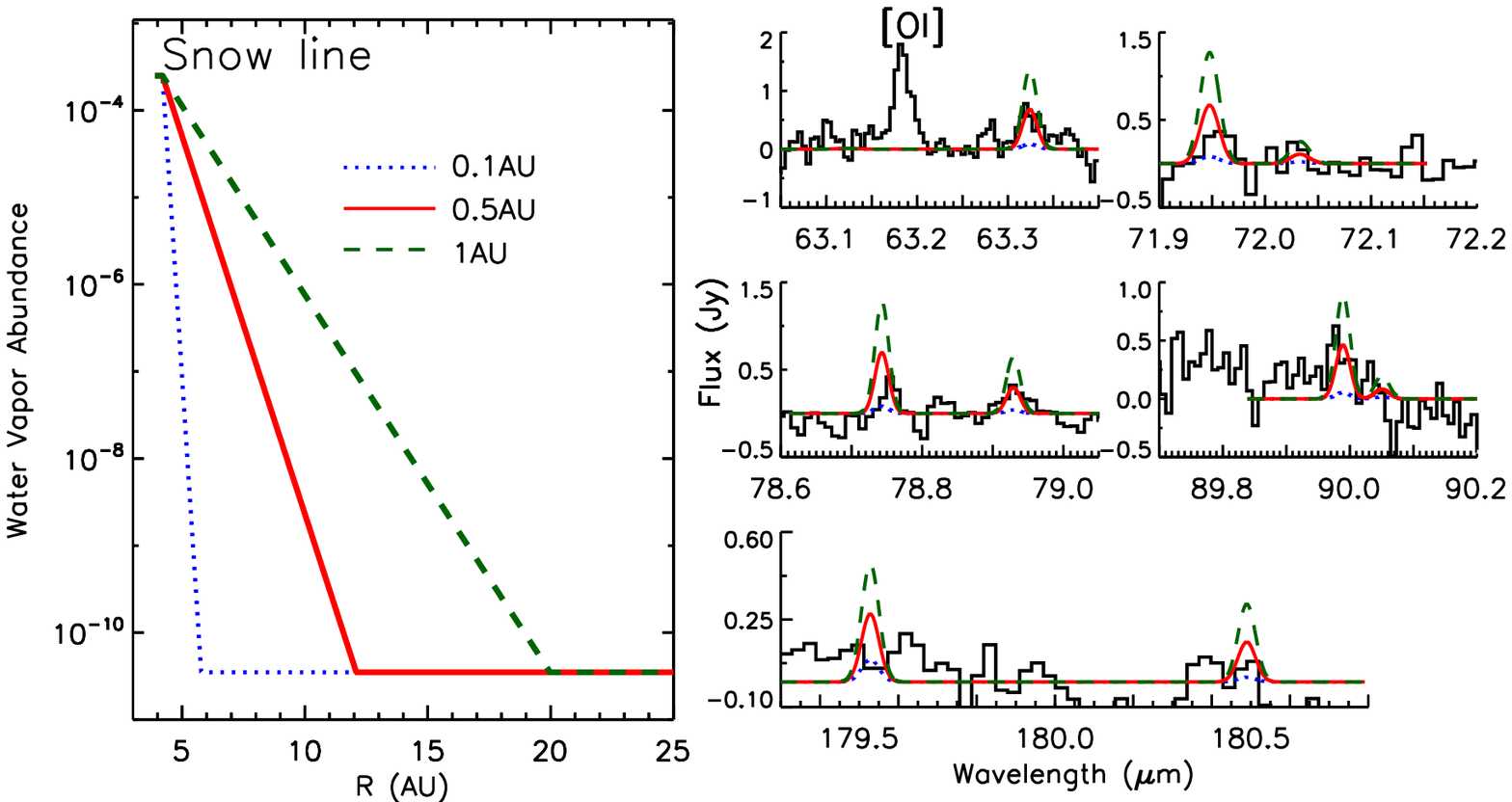}
\caption{Constraints on the sharpness of the snow-line in TW Hya. The left panel shows model water vapor abundance decreases from $\mathrm{X}_{H_2O} \sim 2.5\times10^{-4}$ to $3.5\times10^{-11}$ with three different radial scale lengths:  0.1\,AU (dot line), 0.5\,AU (solid line) and 1\,AU (dash line). The right panel shows the modeled LTE line fluxes over-plotted on \textit{Herschel} PACS spectra. The line styles are the same as in the left panel. }
\label{fig:pacs}
\end{figure*}

Specifically, we modify the abundance step function with an exponential drop-off beyond the snow line:
\begin{equation}
\label{ }
\mathrm{X}_{\rm H_2O}(R>R_{\rm snow}) = (X_{\rm ring}-X_{\rm outer}) e^{{\left( \frac{R_{\rm snow}-R}{R_{\rm eff}}\right)} }+ X_{\rm outer},
\end{equation}

\noindent where $R_{\rm eff}$ is the scale length of the abundance change. We consider $R_{\rm eff}$ = 0.1, 0.5, and 1\,AU, and find  
that models with $R_{\rm eff} \gtrsim $ 0.5\,AU produce more flux than is observed by the {\it Herschel} PACS instrument (Figure~\ref{fig:pacs}). This supports our original assumption of a step function, and we conclude that the surface snow line in TW Hya is radially narrow; that is, it occurs over a region that is a small fraction of its distance to the star.

\subsubsection{The outer disk}
An extremely low vertically averaged abundance beyond the snow line, $X_\mathrm{outer}$=3.5$\times10^{-11}$, is required by  LTE fits to the HIFI ground state water lines at 267 and 537 $\mum$. Indeed, the strongest constraint on the outer disk water vapor abundance is provided by the ground-state lines, although significant limits are also provided by slightly higher-lying transitions in the PACS range, in particular the 179.5\,$\mu$m transition. The low water vapor content of  the outer disk is consistent with the analysis of \citet{Hogerheijde11} and the HIFI non-detection of water lines in another transitional disk, DM Tau \citep{Bergin10}. Specifically, our $X_\mathrm{outer}$ is comparable to the global disk-averaged value derived by \citet{Hogerheijde11} -- 7.3$\times 10^{21}$ g of water in a 1.9$\times 10^{-2}$ M$_{\odot}$ disk, or $X_\mathrm{H_2O}$=2.2$\times10^{-11}$ -- but as these authors stress there is likely significant vertical structure in the water vapor abundance and a large ice reservoir beyond 4-5 AU.

\section{Discussion}
\label{sec:results}

From an analysis of the water emission lines from the TW Hya disk over the $10-567\,\mum$ interval, both warm ($\sim$220 K)  and cold water vapor are found to be present. The warm water emission most likely originates in a narrow ring region between 4--4.2 AU where abundant water vapor carries much of the cosmically available oxygen, $\mathrm{X}_{H_2O} \sim 10^{-4}$. Outside of 4.2 AU, the vertically averaged  water vapor column density decreases dramatically over a radial distance less than 0.5 AU, marking the location of the disk's surface snow line.  Each of the datasets included in this study uniquely constrain one aspect of the distribution profile, and so we demonstrate the importance of the use of multi-wavelength datasets.  Here we discuss some implications of our results.

\subsection{The origin of the surface water vapor}

How does this study inform the origin of the observed water vapor, and of water in protoplanetary disks in general? There are two potential sources of water: One is in situ gas-phase formation, while the other is grain surface formation in a cold reservoir -- which could be the disk itself or the primordial protostellar cloud. In the second case, a mechanism is needed to transport the ice inwards and upwards to a location where it can either thermally evaporate or be photodesorbed and observed in the form of vapor (see, for example, \citealp{Supulver00}).

In the warm inner disk (r$<$1AU, z/r$<$0.2), H$_2$O can be formed in the gas phase \citep{Glassgold09, Bethell09}, even in the presence of substantial photolyzing radiation. The key is a sufficient supply of H$_2$, which drives the formation of water via successive hydrogenation of atomic oxygen in reactions with significant activation barriers. 
More precisely, we find that using the photodissociation rate calculated only at Ly$\alpha$ (which dominates the 3$\times 10^{-3}L_\odot$ FUV continuum flux), and water reaction rates from \citet{Baulch72} -- k$_{\rm OH}$=3.0$\times10^{-14} T e^{-4480/T}\, \mathrm{cm}^3$ molecule$^{-1}$s$^{-1}$ and k$_{\rm H_2O}$=3.6$\times10^{-11} T e^{-2590/T}\,\mathrm{cm}^3$ molecule$^{-1}$s$^{-1}$ -- this simplified chemical model predicts water abundances near 10$^{-4}$ when $T_g\gtrsim$200\,K, $n_g$$>$10$^9$ cm$^{-3}$. These conditions are satisfied in the inner disk and the cavity wall, but not necessarily in the photon-dominated upper layers of the outer disk. 

Thus, production of water via gas-phase reactions can potentially explain the high abundance of $\sim$220\,K water vapor observed inside the TW Hya snow line, even if the initial conditions are atomic, and only until the elemental abundance of oxygen is depleted \citep[$\sim 5\times 10^{-4}$ relative to H, ][]{Frisch03}. Indeed, modern thermo-chemical models \citep{Woitke09, Najita11, Walsh12} predict both that much of the oxygen is locked up in water vapor in the warm inner disk and at large scale heights where the gas is heated by UV and X-ray photons. 

The best fit water content just inside the TW Hya snowline is not sufficiently high to distinguish between in situ gas-phase formation and evaporating icy bodies, but is consistent with thermo-chemical models. Well inside the transition radius, the water abundance drops by two orders of magnitude, in conflict with this simple model. However, the models by Woitke et al. and Walsh et al. do not include the depleted inner region of TW Hya, making direct comparisons difficult for the innermost disk. 

The warm gas-phase chemistry will not operate in the outer disk.
 Here, the primordial surface ice chemistry will dominate \citep{Tielens82}, resulting in a huge reservoir of icy dust and larger bodies from which water molecules must be thermally evaporated or desorbed by high-energy particles and photons.  As discussed in \citet{Bergin10} and \citet{Hogerheijde11}, models of photodesorption from icy grains find that a depletion of icy grains from the outer disk surface, presumably by setting to the midplane, is needed to reproduce the low observed outer disk water vapor abundances.  A zeroth-order expectation is that the water vapor abundance across the snow-line can be treated as a step function \citep{Ciesla06}, with high inner disk water vapor abundances created by a mixture of gas-phase reactions and evaporating icy bodies and a low outer disk abundance maintained by desorption.

We find a very low (vertically averaged) outer disk water abundance of a few times $10^{-11}$ per hydrogen, consistent with previous Herschel-HIFI observations, but in conflict with the static thermo-chemical models, which indicate surface abundances of $10^{-7}-10^{-9}$ per H. We interpret this as further evidence for settling of icy grains in the outer disk of TW Hya.

Another potential test for icy-grain contributions to the observed disk water vapor content would be a measurement of the O/P ratio.  For gas phase formation of water at $>$200\,K, an O/P ratio of 3 is expected, while the O/P ratio for evaporation or sputtering can be considerably less than 3 for grain surface temperatures less than $\sim40$\,K.  However, for higher temperature ices formed in regions close to the snow line, the O/P ratio may still be near 3.  Thus, detailed measurements of the O/P ratio will be needed to trace the origin of the snow-line water vapor,
such as could be obtained, for example, with next generation ground-based (VLT-VISIR) or space-based (JWST-MIRI) thermal-IR spectrographs.

\subsection{The water vapor abundance distribution in transitional disks}
\begin{figure}
\includegraphics[width=6.75cm, angle=90]{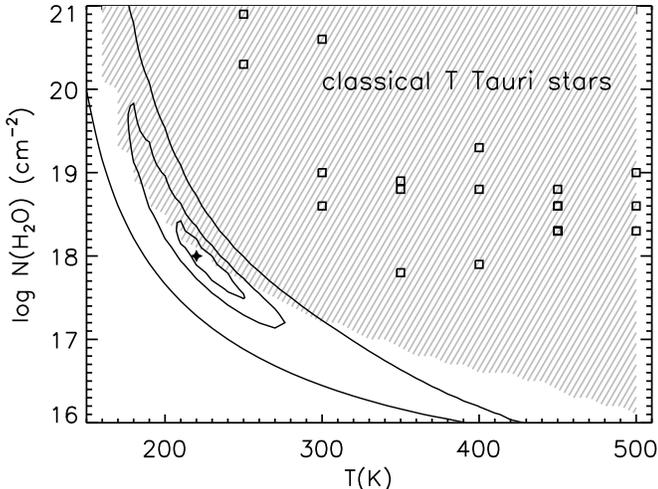}
\caption{The $\chi^2$ surface for LTE slab model water fits for TW Hya as a function of $N$ and $T$, with an effective radius $R$=1.0\,AU. The contour plot is based on \textit{Spitzer} IRS LH spectra, with best-fit parameters of $T$ = 220\,K and $N$=$10^{18}$ cm$^{-2}$. The shadowed area depicts the parameter space that is excluded by the water upper flux limits from the \textit{Spitzer} IRS SH spectrum. Empty squares depict the best LTE slab model fit results for cTTs with water detections in the \textit{Spitzer}-IRS wavelength range \citep{Salyk11}.}
\label{fig:chi2}
\end{figure}

The \textit{Spitzer} IRS spectrum reveals that the water vapor emission from TW Hya is different from that of typical, but less evolved, protoplanetary disks around solar-type stars. A comparison of the basic properties of the TW Hya water emission is shown in Figure \ref{fig:chi2}, where an LTE slab model (disk-averaged) temperature and column density are compared to those of cTTs disks \citep{Salyk11}. This is consistent with the detailed radial abundance structure derived for TW Hya. 

Although the relative contributions to the observed water vapor in TW Hya from gas-phase reactions and evaporating icy planetesimals is unknown, 
it may be the ideal target to observe the evaporation process as its relatively small `hole' size allows the snow line to reside in the optically thick outer disk.  In contrast, many of the known transitional disks have transition radii of 10 AU or greater \citep{Andrews11}, and the snow-line would be located well inside the optically thin region of the disk.  
They also typically lie two or three times more distant than TW Hya.
Therefore, comparisons with additional transitional disks may need to wait for the availability of more sensitive instruments.

In contrast to the $\sim$1 AU snow line found by \citet{Meijerink09} for cTTs disks, the TW Hya snow-line is located at a significantly larger radius, in spite of its lower luminosity as compared to DR Tau and AS 205 N.  In the case of transitional disks, however, the exposure of the transition radius to nearly unobstructed stellar light and the `face on' nature of the cavity wall at the transition radius should heat the disk more readily at large radii and push the snow-line outwards. One interesting consequence of this property is that if giant planets form and clear out a hole in the disk, creating a transitional disk, the subsequent migration of the snow line to larger radii could slow the growth of new planetary cores over a larger range of radii, out to the location of the new snow line.

At the same time, new planetesimal formation may be {\it catalyzed} near the new snow line due to enrichment of water by a combination the radial cold finger effect of \cite{Stevenson88} and inwards radial migration of icy bodies \citep{Ciesla06}. The observed high abundance of water around 4 AU in TW Hya is supportive of a scenario in which new icy planetesimal formation, and perhaps even giant planet formation, is ongoing at this location. 

Another interesting feature of the distribution of water vapor abundance in TW Hya is the relatively dry inner disk. The water vapor abundance in the innermost disk ($<$ 0.5\,AU) is actually unconstrained by the \textit{Spitzer} data due to beam dilution,
but for $r >$0.5 AU the abundance upper limit is two orders of magnitude below that estimated in cTTs disks, and the lack of water in inner disk can be explained by two possible scenarios. 

First, the inner disk really is empty between 0.5-4\,AU, a scenario consistent with near-IR interferometric data.  \citet{Akeson11} found an inner optically thick ring around 0.5 AU plus an outer optically thick disk starting at 4 AU can match the full suite of near-IR interferometric data and the general SED shape of TW Hya.  Recent 8 - 18 $\mu$m speckle imaging, however, has suggested a more continuous dust distribution out to 4 AU and perhaps the presence of a companion \citep{Arnold12}.

 Further, analysis of the ALMA CO $J$=2-1 and 3-2 Science Verification observations of TW Hya has revealed gas signatures down to radii as close as 2 AU \citep{Rosenfeld12}. Thus, the more likely scenario for the structure of the TW Hya disk is that outlined in Figure \ref{fig:column}, namely an inner disk in which CO remains abundant since the kinetic temperature is much higher than that needed for freeze-out ($\sim 20$\,K).Water in the inner disk can still be subjected to some depletion, or in the case that a companion gates the accretion flow into the inner disk any water-ice rich grains that settle to the (outer disk) mid-plane may be blocked from further inward migration. If this is the case, there should be a large difference in the gas phase C/O ratios between the inner and outer parts of the disk. Thus, the quantity of gas along with its C/O content are potentially probes that can distinguish whether a given transition disk has been sculpted by planet formation or other mechanisms (grain growth or photo-evaporation, for example; see \citealt{Najita11} for a further discussion of how changes in the C/O ratio can impact disk chemistry). 

A final aspect of these observations is the short scale length ($<$0.5\,AU) over which the dramatic water vapor decrease occurs, a distance in conflict with scenarios that consider only freeze-out.
Such a radial profile may be further evidence for a vertical cold finger effect, in which the surface snow-line location reflects that at the mid-plane, perhaps due to mixing \citep{Meijerink09}.

\subsection{Molecular mapping with multi-wavelength spectra}
We have demonstrated here how multi-wavelength spectroscopic data can be used to reconstruct molecular abundances, even when the observations are spatially (and often spectrally) unresolved.The snow line differences inferred for the tiny sample observed to date are in accord with the changing physical conditions that passive cTTs disks experience as they evolve. Much more detailed constraints and comparisons to models will become available over the coming years by exploiting the significant infrared database that exists for stars+disks of varying mass and evolutionary state \citep{Pontoppidan10a,Salyk11} along with new observations at longer wavelengths.

Because protoplanetary disks are complicated structures, the multi-wavelength molecular mapping method should be used with some caution, however. The flux of each line will arise from a range of disk regions, with lines at various wavelengths characterizing different radii and/or vertical depths, and an overview of several of the key water vapor tracers for the best fitting TW Hya model is presented in Fig.~\ref{fig:line_region}.  As demonstrated in this work, it can take considerable effort to build realistic models for an individual source, and efforts herein were aided by a large set of ancillary observations of this well-studied disk.  Nevertheless, it is important to stress that relatively fewer uncertainties are introduced in this type of modeling than with the use of full thermal-chemical disk models, in which chemical abundances are sensitive to a wide array of model parameters, including FUV and X-ray fluxes, accretion rate, disk viscosities and transport rates, dust-versus-gas settling geometries, grain-surface reaction rates, and so on \citep[e.g.][]{Heinzeller11}.   Further, studies of large disk samples will require fast, robust models. We therefore advocate for the direct measurement of chemical abundances with relatively few free parameters, such as is described in this work, to be used in conjunction with physical intuition derived from the more complex thermal-chemical models.  

It is interesting to note that the precision of derived abundance profiles is primarily determined by the availability of high-quality spectra across a wide range of excitation energies, as well as by the knowledge of the overall disk structure, especially the gas temperature.  Thus, more precise measurements can be obtained as spectrographs with higher spectral resolution and sensitivity become available, irrespective of the instrumental spatial resolution (provided the disk structure is reasonably well characterized) -- {though of course instruments that provide detailed kinematic profiles of lines with a significant range of excitation energies will provide the most stringent probes. This technique thus provides an ideal means to study chemical structure at the size scales relevant to planet formation.  

With a properly chosen spectral suite, we have demonstrated that it is possible to probe molecular abundance distributions on AU scales, and it is worth emphasizing that this method is highly complementary to the capabilities of ALMA. At its longest baselines (16\,km), the full ALMA will be able to resolve a nearby disk (140\,pc) on AU scales at its shortest operational wavelength, $\sim$400\,$\mum$, in dust emission. However, due to the quantum-limited nature of heterodyne receivers and the available collecting area, it will be a severe challenge even for ALMA to robustly image the warm molecular gas inside of 10\,AU -- the principle formation region of terrestrial and giant planets according to core accretion theory. Thus, for the foreseeable future, multi-wavelength methods (which can incorporate ALMA spectral data cubes) offer perhaps the best means of deriving the molecular abundance patterns in planet-forming environments.  Table \ref{tab:tracer} presents the general selectivity of each wavelength window for disks around Sun-like stars. 

Although current observational S/N levels limit chemical abundance studies to a few species, this technique can in principle be applied to the study of any molecule whose transitions cover sufficient excitation space.  The gas phase distribution of condensible species provides particularly valuable information on the temperature, density, chemical and transport patterns in disks; and used together, observations of a wide variety of molecules are best suited to the isolation and thus understanding of particular disk processes. To combine both infrared and (sub)mm data on the inner/outer disk, polar species with greatly differing sublimation temperatures are preferred. The most obvious candidates, beyond water vapor and OH, are CO and HCN, which are widely detected in surveys of protoplanetary disks. The former provides a fiducial chemical marker that only becomes significantly affected by freeze-out at very low temperatures, while the abundance drop for HCN across the snow line may well be sensitive to the C/O ratio in the gas.

\begin{deluxetable}{lll}
\tablecolumns{3}
\tablewidth{0pt} 
\tablecaption{Volatile line tracers at different disk radii}
\tablehead{
\colhead{Wavelength}     & \colhead{Radius}  &\colhead{Facility} \\
\colhead{$\mum$}                      & \colhead{AU}        &\colhead{}
}
\startdata
2-5  & 0.1 &NIRSPEC, CRIRES\\
6-10 & 0.1- 1 & SOFIA FORECAST \\
10-35& 0.5- 5 & Spitzer IRS \\
40 - 200 & 1-100 & Herschel PACS\\
200- 30000 & $>$50 & HIFI, ALMA
\enddata
\label{tab:tracer}
\end{deluxetable} 
 
\begin{figure}
\includegraphics[width=9cm]{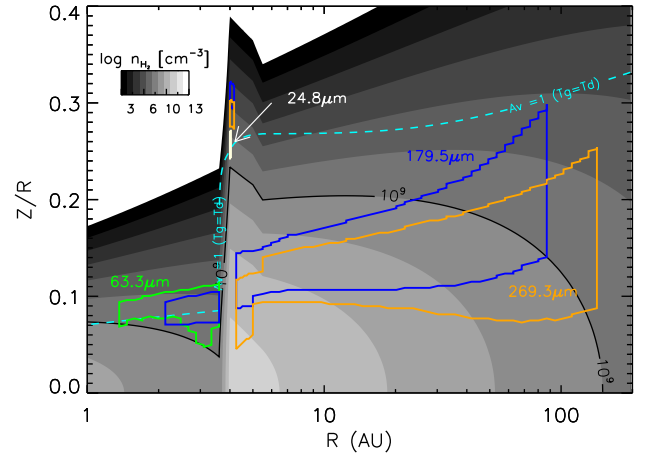}

\vspace{0.0001in}
\caption{ The radial and vertical locations bounding water vapor emission from the TW Hya disk. The boxes mark the 15\% and 85\% cumulative radial line flux limits (vertical lines) and the heights where 15\% and 85\% of the line flux arises from each vertical column (horizontal lines). The gas density of the disk is shown in greyscale, and the location of the $A_v$$\sim$1 gas/dust decoupling transition ($T_g = T_d$) is shown by the dotted line. As a guide to the eye, the $n_{H_2} = 10^9$ cm$^{-3}$ density contours are labeled by the solid line. Excitation and dust/line opacity are principally responsible for the lack of sensitivity to the cold, dense disk midplane (c.f. Figure \ref{fig:SED}).}
\label{fig:line_region}
\end{figure}

\section{Conclusions}

This paper uses multi-wavelength spectra to probe the water vapor distribution in protoplanetary disks, and demonstrates the first application of this method to investigate the water vapor abundance from 0.5-200\,AU in the transitional disk TW Hya. Our modeling shows that there is a narrow region between 4-4.2\,AU where water vapor is warm ($\sim$220\,K) and optically thick, resulting in a high abundance ($\mathrm{X}_{H_2O}$$\sim$10$^{-4}$). Outside the snow-line, the water vapor column density in the disk atmosphere decreases dramatically over a scale length of less than 0.5\,AU due to freeze out, resulting in a vertically integrated vapor abundance of $\mathrm{X}_{H_2O}$$\sim$10$^{-11}$ at large radii.

Both current and near-term astronomical facilities from the near-IR to radio frequencies can generate a rich molecular data set of the emission lines from protoplanetary disks. We expect the multi-wavelength molecular mapping method will soon be applied to a substantial ensemble of disks and molecular species.

\section{Acknowledgements}

This work is based in part on observations made with the \textit{Spitzer} Space Telescope, which is operated by the Jet Propulsion Laboratory, California Institute of Technology under a contract with NASA. \textit{Herschel} is an ESA space observatory with science instruments provided by European-led Principal Investigator consortia and with important participation from NASA. Funding for this work (to G. Blake) was provided by the NASA Astrobiology/Origins of Solar Systems programs and NSF Astronomy \& Astrophysics. C. Salyk gratefully acknowledges NOAO Leo Goldberg Postdoctoral Fellowship support. We also thank the anonymous referee for a thorough and insightful report, which helped to improve this paper.

\bibliographystyle{apj}
\bibliography{ms}Ä

\end{document}